\documentclass[preprint,12pt]{elsarticle}

\usepackage{graphics}

\usepackage{mathrsfs}
\usepackage{amsmath}
\def\Vec#1{\mbox{\boldmath $#1$}}
\newcommand{\Dd}{D_{\!\text{d}}}

\journal{Physica A}

\begin{document}

\begin{frontmatter}



\title{Irreversibility and Entropy Production \\ in Transport Phenomena I}


\author{Masuo Suzuki}
\address{Tokyo University of Science\\Kagurazaka 1-3, Shinjuku, Tokyo, 162-8601}
\ead{msuzuki@rs.kagu.tus.ac.jp}

\begin{abstract}

The linear response framework was established a half-century ago, but no persuasive direct derivation of entropy production has been given in this scheme.
This long-term puzzle has now been solved in the present paper.
The irreversible part of the entropy production in the present theory is given by $(dS/dt)_{\text{irr}}=(dU/dt)/T$ with the internal energy $U(t)$ of the relevant system.
Here, $U(t)=\langle \mathcal{H}_0\rangle _t={\text{Tr}}\mathcal{H}_0 \rho (t)$ for the Hamiltonian $\mathcal{H}_0$ in the absence of an external force and for the density matrix $\rho(t)$. 
As is well known, we have $(dS/dt)_{\text{irr}}=0$ if we use the linear-order density matrix $\rho _{\text{lr}}(t)=\rho_0+\rho_1(t)$.
Surprisingly, the correct entropy production is given by the {\itshape{second-order symmetric}} term $\rho_2(t)$ as $(dS/dt)_{\text{irr}}=(1/T){\text{Tr}}\mathcal{H}_0 \rho'_2(t)$.
This is shown to agree with the ordinary expression $\Vec{J}\cdot\Vec{E}/T=\sigma E^2/T$ in the case of electric conduction for a static electric field $\Vec{E}$, using the relations $\text{Tr}\mathcal{H}_0\rho'_2(t)=-{\text{Tr}}\dot{\mathcal{H}}_1(t)\rho_1(t)={\text{Tr}}\Vec{\dot{A}}\cdot\Vec{E}\rho_1(t)=\Vec{J}\cdot\Vec{E}$ (Joule heat), which are derived from the {\itshape{second-order}} von Neumann equation $i\hbar d\rho_2(t)/dt=[\mathcal{H}_0,\rho_2(t)]+[\mathcal{H}_1(t),\rho_1(t)]$. Here 
$\mathcal{H}_1(t)$ denotes the partial Hamiltonian due to the external force such as $\mathcal{H}_1(t)=-e\sum_{j}\Vec{r}_i\cdot\Vec{E}\equiv -\Vec{A}\cdot\Vec{E}$ in electric conduction.
Thus, the linear response scheme is not closed within the first order of an external force, in order to manifest the irreversibility of transport phenomena.
New schemes of steady states are also presented by introducing relaxation-type (symmetry-separated) von Neumann equations. The concept of stationary temperature $T_{\text{st}}$ is introduced, which is a function of the relaxation time $\tau_{\text{r}}$ characterizing the rate of extracting heat outside from the system.
The entropy production in this steady state depends on the relaxation time.
A dynamical-derivative representation method to reveal the irreversibility of steady states is also proposed.
The present derivation of entropy production is directly based on the first principles of using the projected density matrix $\rho_2(t)$ or more generally symmetric density matrix $\rho_{\text{sym}}(t)$, while the previous standard argument is due to the thermodynamic energy balance.
This new derivation clarifies conceptually the physics of irreversibility in transport phenomena, using the symmetry of non-equilibrium states, and this manifests the duality of current and entropy production.
\end{abstract}


\end{frontmatter}

\section{Introduction}
\label{intro}
The basic problems of irreversibility and transport phenomena have been discussed for many years from many viewpoints[1-12].
Boltzmann's $H$-theorem shows the first attempt to explain the broken symmetry of entropy change in time, namely irreversibility, even though it is based on stochastic equations (such as the Boltzmann equation).
On the other hand, transport phenomena were studied phenomenologically first by Einstein in 1905, and later by Onsager\cite{13,14}.
Green\cite{3} formulated transport coefficients using stochastic equations including the time change of microscopic fluctuations\cite{5}.
Kubo and Tomita\cite{8} succeeded, for the first time, in formulating magnetic linear responses on the basis of the von Neumann equation.
Nakano\cite{9} applied Kubo-Tomita's method to the problem of electric conduction, though the two problems had been regarded to be different in the sense that the former is reduced to an equilibrium problem for zero frequency limit, but the latter is still in non-equilibrium even for the static limit\cite{10}.
Kubo\cite{4} established a general theory of linear responses including the magnetic response and electric conduction, as is briefly reviewed in Section 2.
Previous arguments on the energy dissipation and entropy production in the linear response scheme are summarized in Section 3.
First-principles derivation of entropy production due to the symmetric part of the density matrix is given in Section 4, in order to clarify conceptually irreversibility in transport phenomena as an example of static electric conduction.
In Section 5, new schemes of steady states and entropy production are presented by introducing relaxation-type von Neumann equations.
In Section 6, the dynamical-derivative representation method is proposed to derive the entropy production of steady states.
A summary and discussion are given in Section 7.

\section{Brief review of the linear response scheme with reference to irreversibility}

As in Kubo's paper\cite{4}, we start with the following von Neumann equation for the density matrix $\rho(t)$:
\begin{equation}
i\hbar \frac{\partial }{\partial t}\rho(t)=[\mathcal{H}(t),\rho(t)], \label{eq:1}
\end{equation}
where the total Hamiltonian is given by 
\begin{equation}
\mathcal{H}(t)=\mathcal{H}_0+\mathcal{H}_1(t);\quad\mathcal{H}_1(t)=-\Vec{A}\cdot\Vec{F}(t).\label{eq:2}
\end{equation}
For the physical meaning of $\mathcal{H}_0$, see Kubo's paper\cite{4} and Section 4 (below Eq.(\ref{eq:25})).
Even the conductivity (\ref{eq:10}) is expressed only by $\mathcal{H}_0$ as well as the current operator $\Vec{j}$.
Now we expand $\rho(t)$ as
\begin{equation}
\rho_{\text{lr}}(t)=\rho_0+\rho_1(t);\quad\rho_0=e^{-\beta\mathcal{H}_0}/Z_0(\beta)\label{eq:3}
\end{equation}
with $Z_0(\beta)={\text{Tr}}\exp(-\beta\mathcal{H}_0)$, up to the first order of an external force $\Vec{F}$.
Then, we have
\begin{equation}
i\hbar\frac{\partial}{\partial t}\rho_1(t)=[\mathcal{H}_0,\rho_1(t)]+[\mathcal{H}_1(t),\rho_0].\label{eq:4}
\end{equation}
The solution of this equation is given by
\begin{equation}
\rho_1(t)=\int_{t_0}^{t}\mathcal{U}_0(t-s)\frac{1}{i\hbar}[\mathcal{H}_1(s),\rho_0]\mathcal{U}^{\dagger}_0(t-s)ds\label{eq:5}
\end{equation}
with the unitary operator $\mathcal{U}_0(t-s)$ defined by
\begin{equation}
\mathcal{U}_0(t-s)=\exp\left(\frac{t-s}{i\hbar}\mathcal{H}_0\right).\label{eq:6}
\end{equation}
Using Kubo's identity\cite{4} or the quantum derivative\cite{15}
\begin{align}
\left[A,e^{-\beta \mathcal{H}_0}\right]&= e^{-\beta \mathcal{H}_0}\int_{0}^{\beta}e^{\lambda \mathcal{H}_0}[\mathcal{H}_0,A]e^{-\lambda \mathcal{H}_0}d\lambda \notag
\\
&=-i\hbar e^{-\beta \mathcal{H}_0}\int_{0}^{\beta}e^{\lambda \mathcal{H}_0}\dot{A}e^{-\lambda \mathcal{H}_0}d\lambda \notag
\\
&=-i\hbar e^{-\beta \mathcal{H}_0}\int_{0}^{\beta}\dot{A}(-i\hbar \lambda)d\lambda,\label{eq:7}
\end{align}
the first-order density matrix $\rho_1(t)$ is expressed as
\begin{equation}
\rho_1(t)=\rho_0\int_{t_0}^{t}ds\int_{0}^{\beta}d\lambda \Vec{F}(s)\cdot\Vec{\dot{A}}(s-t-i\hbar\lambda), \label{eq:8}
\end{equation}
where $\Vec{\dot{A}}$ denotes the current $\Vec{j}$ of $\Vec{A}$, namely $\Vec{j}=\Vec{\dot{A}}=[\Vec{A},\mathcal{H}_0]/i\hbar$.
Finally, we arrive at the well-known formula\cite{4,9}
\begin{subequations}
\begin{align}
\Vec{J}\equiv \langle \Vec{j}\rangle _t &= {\text{Tr}}\left\{ \left( \rho_0+\rho_1(t) \right) \Vec{j} \right\} \notag
\\
&= \int_{t_0}^{t}ds\int_{0}^{\beta}d\lambda \langle \Vec{j}\Vec{j}(t-s+i\hbar \lambda )\rangle _0 \Vec{F}(s),\label{eq:9a}
\end{align}
which is reduced to
\begin{equation}
\Vec{J}= \int_{0}^{\infty}ds\int_{0}^{\beta}d\lambda e^{-\epsilon s} \langle  \Vec{j}\Vec{j}(s+i\hbar \lambda) \rangle _0 \Vec{F}(t-s)\label{eq:9b}
\end{equation}
\end{subequations}
by taking the limit $t_0\rightarrow -\infty$ after the thermodynamic limit.
The order of the above two limiting procedures and the adiabatic factor $e^{-\epsilon s}$ are vital to assure the non-vanishing value of the current, namely to realize the irreversibility in transport phenomena, as will be shown in Section 4.
In particular, for the static electric conduction $\Vec{F}(t)=\Vec{E}$(constant), we have
\begin{equation}
\Vec{J}=\sigma\Vec{E};\quad\sigma=\int_{0}^{\infty}dt\int_{0}^{\beta}d\lambda e^{-\epsilon t} \langle  \Vec{j}\Vec{j}(t+i\hbar \lambda) \rangle _0 \label{eq:10}
\end{equation} 
with the current $\Vec{j}=\Vec{\dot{A}}$ and $\Vec{A}=e\sum_{j}\Vec{r}_j$.
The finiteness ($\sigma >0$) of this expression is important to discuss later the entropy production and irreversibility in electric conduction from the first-principles based on the von Neumann equation.
Note that this condition of finiteness ($\sigma >0$) is violated in the opposite limit $t_0\rightarrow -\infty$ for a finite volume in Eq.(\ref{eq:9a}), because we obtain $\sigma\propto \text{Tr}A[A,\rho_0]=0$ by integrating $\Vec{\dot{A}}(t)\equiv \Vec{j}(t)$ in Eq.(\ref{eq:9a}) with Eq.(\ref{eq:5}) with respect to time $t$ in this unphysical condition.

\section{Previous arguments on the energy dissipation and entropy production in the linear response scheme}

The problem of entropy production is very subtle even in the linear response scheme.
The first puzzle is that the ordinary formula $S(t)=-k_{\text{B}}\text{Tr}\rho(t)\log\rho(t)$ is useless for our purpose, because it is time-independent, as is well known.
There have been many arguments proposed on the energy dissipation and entropy production in the linear response scheme.

Some typical arguments are briefly reviewed in order to clarify how insufficient they are from the present new point of view.

\subsection{Thermodynamic derivation of energy dissipation based on energy balance in a steady state}

The energy conservation law yields the following balance equation of the change of the internal energy $dU$ in the form
\begin{equation}
dU=\Delta Q+\Delta W,
\end{equation} 
where $\Delta Q$ is the heat transferred to the system during the process of change, and $\Delta W(=\Vec{J}\cdot\Vec{E}\text{ per unit time})$ denotes the work done by the force (electric field $\Vec{E}$) on the system.
The internal energy is assumed to be constant, namely $dU=0$ to keep the temperature of the system constant.
Note that 
\begin{equation}
U'_{\text{lr}}(t)\equiv \frac{d}{dt}\text{Tr}\mathcal{H}_0\rho_{\text{lr}}(t)=\text{Tr}\mathcal{H}_0\rho'_1(t)=0\label{eq:11}
\end{equation}
using the expression (\ref{eq:8}).
In this sense, at a glance the above proposition $dU=0$ seems to be appropriate.
Then, we have $-\Delta Q=\Delta W=\Vec{J}\cdot\Vec{E}=\sigma E^2$.
This is the so-called Joule heat $|\Delta Q|$, or energy dissipation transferred to the heat reservoir.
This argument seems to be natural, because the amount of the dissipated energy $|\Delta Q|=\sigma E^2$ is correct.
However, in order to realize the above situation in which the system is stationary, namely the temperature $T$ is constant, we have to modify the von Neumann equation (\ref{eq:1}) by introducing an appropriate interaction with the heat reservoir and boundary conditions.
This modification is not so simple, as has been tried by many people.

If we avoid such complication of modification, then we are able to accomplish our goal to understand clearly the mechanism or essence of irreversibility in transport phenomena, namely to derive directly the entropy production $(dS/dt)_{\text{irr}}=U'(t)/T(t)$ from the first-principles using the von Neumann equation (\ref{eq:1}).
This will be presented in the succeeding section.

\subsection{Primitive try to derive entropy production from $S(t)=-k_{\text{\rm{B}}}{\text{\rm{Tr}}}\rho(t)\log\rho(t)$ approximating $\rho(t)$ by the linear density matrix $\rho_{\text{\rm{lr}}}(t)$}

One expects to derive a correct expression of entropy change from
\begin{equation}
S_{\text{lr}}(t)=-k_{\text{B}}\text{Tr}(\rho_0 + \rho_{1}(t))\log (\rho_0 + \rho_{1}(t)),\label{eq:13}
\end{equation}
but the time derivative of this entropy calculated up to the second order of the electric field $\Vec{E}$ is not equal to $\sigma E^2/T$.
The reason of this discrepancy will become clear in the succeeding section.
It should also be noted that the expression (\ref{eq:13}) depends on time $t$, while the expression $S(t)=-k_{\text{B}}\text{Tr}\rho(t)\log\rho(t)$ does not as mentioned before.
(The introduction of approximation (in the first order of $\Vec{E}$) is expected to yield irreversibility, for it sometimes corresponds to the contraction of information.)

\subsection{Zubarev-type argument with a "negative sign" problem}

Zubarev extended the linear response scheme to a non-linear regime by introducing a non-equilibrium statistical operator of the form\cite{11}
\begin{align}
\rho_{\text{Zub}}(t)&=\exp \left(-\Phi (t) -\beta \mathcal{H}_0 +\beta\int_{t_0}^{t}ds e^{\epsilon (s-t)}\Vec{j}(s-t)\cdot \Vec{E} \right)\notag
\\
&\simeq \rho_{\text{lr}}(t)+\cdots =\rho_0 + \rho_1(t)+\cdots
\end{align}
in the case of static electric conduction. 
(Zubarev discussed a general case.)
That is, $\rho_{\text{Zub}}(t)$ agrees with $\rho_{\text{lr}}(t)=\rho_0 + \rho_1(t)$ as far as the linear term is concerned, and conversely the former can be constructed so that the expanded series of the exponential part of it may agree with $\rho_{\text{lr}}(t)$ up to first order.

Zubarev defined the entropy $S_{\text{Zub}}(t)$ of this system by 
\begin{equation}
S_{\text{Zub}}(t)=-k_{\text{B}}\text{Tr}\rho_{\text{Zub}}(t)\log\rho_{\text{loc}},
\end{equation}
where
\begin{equation}
\rho_{\text{loc}}=e^{-\beta(\mathcal{H}_0+\mathcal{H}_1)}/Z(\beta);\quad Z(\beta)=\text{Tr}e^{-\beta(\mathcal{H}_0+\mathcal{H}_1)}\label{eq:16}
\end{equation}
in the present problem with $\mathcal{H}_1=-\Vec{A}\cdot\Vec{E}$ in static electric conduction.
Then, his entropy production is expressed as
\begin{equation}
\frac{d}{dt}S_{\text{Zub}}^{\text{(lr)}}(t)=\frac{1}{T}\text{Tr}(\mathcal{H}_0+\mathcal{H}_1)\rho'_{\text{Zub}}(t). \label{eq:17}
\end{equation}
In the limit of the linear response scheme, this is reduced to
\begin{equation}
\frac{d}{dt}S_{\text{Zub}}^{\text{(lr)}}(t)=\frac{1}{T}\text{Tr}\mathcal{H}_1\rho'_1(t), \label{eq:18}
\end{equation}
because of the different symmetries of $\mathcal{H}_0$ and $\rho_1(t)$.
Thus, Zubarev erroneously concluded that the entropy production (\ref{eq:18}) is given by $\sigma E^2/T(>0)$, at least, in the linear response regime.
However, as will be seen from the third last expression of the right-hand side of Eq.(\ref{eq:33}) in the succeeding section, the correct calculation of Eq.(\ref{eq:18}) yields the negative value $-\sigma E^2/T$.
The discrepancy of  the above two results may come from the following situation.
In the standard treatment of the non-equilibrium average $\langle Q\rangle _t=\text{Tr}Q\rho(t)$ for any observable $Q$, the time derivative of it is expressed in the following two ways:
\begin{equation}
\frac{d}{dt}\langle Q\rangle _t=\text{Tr}Q\rho'(t)=\text{Tr}\frac{dQ}{dt}\rho(t),
\end{equation}
where $dQ/dt$ is given by
\begin{equation}
\frac{dQ}{dt}=\frac{1}{i\hbar}[Q,\mathcal{H}(t)]=-\frac{1}{i\hbar}[\mathcal{H}(t),Q]\label{eq:20}
\end{equation}
with $\mathcal{H}(t)=\mathcal{H}_0+\mathcal{H}_1(t)$.
(This derivative of $Q$ will be denoted later as $\tilde{\Dd}Q$ in Eq.(\ref{eq:61}).)
On the other hand, such an operator $f(\rho(t))$ as the "entropy operator" $\log \rho(t)$ obeys the von Neumann -like equation
\begin{equation}
i\hbar\frac{\partial }{\partial t}f(\rho (t)) =[\mathcal{H}(t),f(\rho (t))],\label{eq:21}
\end{equation}
as is easily proved for any analytic function $f(x)$\cite{11,15}.
If the time derivative of the operator $\log \rho_{\text{loc}}$ is interpreted incorrectly as in Eq.(\ref{eq:21}), an incorrect sign is realized to appear in the calculation, favorable to Zubarev's theory.

\subsection{Relative-entropy formulation}

The relative entropy $S_{\text{rel}}(t)$ defined by
\begin{equation}
S_{\text{rel}}=-k_{\text{B}}\text{Tr}\rho(t) \left(\log\rho(t)- \log \rho_{\text{loc}}\right)\label{eq:22}
\end{equation}
will be often used to discuss the entropy change.
This yields
\begin{equation}
\frac{d}{dt}S_{\text{rel}}(t)=k_{\text{B}}\frac{d}{dt}\left(\text{Tr}\rho(t)\log\rho_{\text{loc}}\right).\label{eq:23}
\end{equation}
Then, we obtain
\begin{equation}
\frac{d}{dt}S_{\text{rel}}^{\text{(lr)}}(t)=-\frac{1}{T}\text{Tr}\mathcal{H}_1\rho'_1(t)\label{eq:24}
\end{equation}
in the linear response scheme.
This has a correct sign in contrast to Eq.(\ref{eq:18}).
However, this usage of the definition (\ref{eq:22}) is regarded to be only a trick to change the wrong sign to the correct one, from our new point of view!
There is no physical justification for utilizing the relative entropy (\ref{eq:22}) in order to discuss the irreversibility, namely the entropy production.
In fact, Eq.(\ref{eq:23}) always vanishes, if we include higher-order terms of $\rho(t)$ as will be seen in Section 7.

\subsection{Variational theory of the entropy production}

Onsagar and Machlup\cite{14} discussed phenomenologically linear transport phenomena, using the variational theory concerning the bilinear form of the energy dissipation with respect to the current.
This phenomenological theory is, from the beginning, based on the correct expression of the entropy production.
Later, Nakano\cite{10} developed a variational theory of energy dissipation based on his microscopic expression of conductivity, by extending Schwinger's theory\cite{16}.

\section{First-principle derivation of entropy production in the linear and non-linear transport phenomena by symmetry separation}

In order to clarify the essence of the irreversibility, namely entropy production in transport phenomena, we concentrate our arguments here on a non-stationary process in static electric conduction described by the Hamiltonian
\begin{equation}
\mathcal{H}=\mathcal{H}_0+\mathcal{H}_1;\quad \mathcal{H}_1=-\Vec{A}\cdot\Vec{E}{\text{ and }}\Vec{A}=e\sum_{j}\Vec{r}_j.\label{eq:25}
\end{equation}

Here, $\mathcal{H}_0$ denotes the Hamiltonian of electrons together with impurity potentials to scatter electrons or with electron-phonon interactions.
Thus, $\mathcal{H}_0$ is assumed here to be {\itshape{symmetric}} with respect to the inversion of space ($\Vec{r}\rightarrow -\Vec{r}$ or spin inversion $\Vec{s}_j\rightarrow -\Vec{s}_j$ in the case of magnetic responses), while $\mathcal{H}_1$ is antisymmetric.
Therefore, the average $\langle \mathcal{H}_0\rangle _t\equiv {\text{Tr}}\mathcal{H}_0\rho(t)$ can be regarded to be such an internal energy $U(t)$ as is identified as the heat energy of the relevant system, for the initial state $\rho_0$ is given by the equilibrium density matrix without an electric field.
Such a Hamiltonian $\mathcal{H}'_0$ as stores electrons in a static way inside (or at the surface) of the system may be added to $\mathcal{H}_0$.
(Note that $\mathcal{H}'_0$ is not a time derivative of $\mathcal{H}_0$.)
However, this may have a symmetry different from $\mathcal{H}_0$, and it is regarded not to contribute to the motion of electrons and consequently the time derivative $\Vec{\dot{A}}$ is still given only by $\mathcal{H}_0$ as $\Vec{\dot{A}}=[\Vec{A},\mathcal{H}_0]/i\hbar$ without using $\mathcal{H}'_0$, and $\mathcal{H}'_0$ is also assumed to be commutable with $\mathcal{H}_0$.
Then, our following arguments still hold.
(Only the energy balance changes: The (input) electric power becomes equal to the sum of the Joule heat ($\Vec{J}_E\cdot\Vec{E}$) and $d\langle \mathcal{H}'_0\rangle _t/dt$.)
These yield a finite (non-zero) value of the conductivity $\sigma$.
This finiteness of $\sigma$ gives the positive entropy production.
See Appendices A and B for intuitive arguments of the role of $\mathcal{H}_0$ in our theory of irreversibility and entropy production.
The vanishing $\sigma$ in a random quantum system with impurity potentials corresponds to Anderson localization. (See ref.\cite{17})

Physically, the internal energy $U(t)$ of the system increases owing to the applied electric field (or power), that is, the system heats up in the present non-stationary situation.
Thus, the temperature $T$ of the system depends on time $t$ as $T=T(t)$.
(This time dependence will be defined later.)
This is the essence of irreversibility in the electric conduction phenomenon.
Thermodynamically, the entropy change $dS$ is given by $dS=dU/T$ using the internal energy change $dU$.
Correspondingly, the entropy production $(dS/dt)_{\!\text{irr}}$ intrinsically produced inside the system by the generation of heat is defined, in our microscopic formulation, by
\begin{equation}
\left(\frac{dS}{dt}\right)_{\!\!\!\text{irr}}=\frac{1}{T}\frac{dU(t)}{dt};\quad U(t)=\text{Tr} \mathcal{H}(t) P_{\text{sym}}\rho(t)={\text{Tr}}(P_{\text{sym}}\mathcal{H}(t)) \rho(t) =\text{Tr}\mathcal{H}_0 \rho(t),\label{eq:26}
\end{equation}
where $P_{\text{sym}}$ denotes the projection operator to the symmetric (or even) part of the relevant operator with respect to the inversion of space or the external field $E$.
In fact, the entropy should be independent of the sign of $E$.
Therefore, it should be an even function of $E$.
Thus, it is described by $\rho_{\text{sym}}(t)$ defined by Eq.(\ref{eq:36b}). 
The definition of $U(t)$ in terms of Eq.(\ref{eq:26}) (not defined by the total Hamiltonian) is essential for the following argument.
It is also instructive to rewrite Eq.(\ref{eq:26}) in the form
\begin{equation}
\left(\frac{dS}{dt}\right)_{\!\!\text{irr}}=-k_{\text{B}}\frac{d}{dt}\text{Tr}\rho(t)P_{\text{sym}}(\log \rho_{\text{loc}}),\label{eq:27}
\end{equation}
using $\rho_{\text{loc}}$ given by Eq.(\ref{eq:16}).

As mentioned in the preceding section, the linear-response density matrix $\rho_{\text{lr}}(t)$ given by (\ref{eq:3}) and (\ref{eq:8}) does not give a finite contribution to the entropy production (\ref{eq:26}), because of the difference of symmetries between $\mathcal{H}_0$ and $\mathcal{H}_1$ (or $\rho_1(t)$).
In this situation, we {\itshape{dare to try}} to study the contribution of the {\itshape{second-order}} term $\rho_2(t)$ to (\ref{eq:26}).

In general, we expand $\rho(t)$ as
\begin{equation}
\rho(t)=\rho_0 +\rho_1(t)+\rho_2(t)+\cdots+\rho_n(t)+\cdots. \label{eq:28}
\end{equation}
From the von Neumann equation (\ref{eq:1}), we obtain the following hierarchical equations
\begin{align}
i\hbar\frac{\partial}{\partial t}\rho_1 (t)&=[\mathcal{H}_0,\rho_1 (t)]+[\mathcal{H}_1(t),\rho_0],\notag
\\
i\hbar\frac{\partial}{\partial t}\rho_2 (t)&=[\mathcal{H}_0,\rho_2 (t)]+[\mathcal{H}_1(t),\rho_1(t)],\notag
\\
\cdots &\cdots\cdots\notag
\\
i\hbar\frac{\partial}{\partial t}\rho_n (t)&=[\mathcal{H}_0,\rho_n (t)]+[\mathcal{H}_1(t),\rho_{n-1}(t)].\label{eq:29}
\end{align}
A solution of Eq.(\ref{eq:29}) is given in the form
\begin{equation}
\rho_n(t)=\mathcal{F}_t\left(\rho_{n-1}(t) \right)=\mathcal{F}_t^{n-1}\left(\rho_1(t) \right)=\mathcal{F}_t^{n}\left(\rho_0 \right),\label{eq:30}
\end{equation}
where
\begin{equation}
\mathcal{F}_t(Q(t))\equiv \frac{1}{i\hbar}\int_{t_0}^{t}\mathcal{U}_0(t-s)[\mathcal{H}_1(s),Q(s)]\mathcal{U}_0^{\dagger}(t-s).\label{eq:31}
\end{equation}
In particular, we have
\begin{align}
&\rho_2(t)=\mathcal{F}_t\left(\rho_1(t)\right)=\mathcal{F}_t^2\left(\rho_0\right)\notag
\\
&=\left(\frac{1}{i\hbar}\right)^{\!\!2}\!\!\!\int_{t_0}^{t}\!\!\!\!ds\!\!\!\int_{t_0}^{s}\!\!\!\!ds' \mathcal{U}_0(t-s)[\mathcal{H}_1(s),\mathcal{U}_0(s-s')[\mathcal{H}_1(s'),\rho_0]\mathcal{U}_0^{\dagger}(s-s')]\mathcal{U}_0^{\dagger}(t-s).\label{eq:32}
\end{align}
Now, we calculate the change of the internal energy, $U'(t)$, using Eq.(\ref{eq:32}).
However, this is rather complicated.
We find that it is much easier to make use of the algebraic equation of the derivative $\rho'_2(t)$, namely Eq.(\ref{eq:29}).
Thus, we arrive finally at
\begin{align}
\left(\frac{dS(t)}{dt}\right)_{\!\!\!\text{irr}}&=\frac{1}{T}\frac{dU(t)}{dt}=\frac{1}{T}\frac{d\langle \mathcal{H}_0\rangle _t}{dt}=\frac{1}{T}\text{Tr}\mathcal{H}_0\rho'_2(t)\notag
\\
&=-\frac{1}{T}\text{Tr}\dot{\mathcal{H}}_1(t)\rho_1(t)=\frac{1}{T}\text{Tr}\Vec{\dot{A}}\cdot\Vec{E}\rho_1(t)\notag
\\
&=\frac{\sigma E^2}{T}>0,\label{eq:33}
\end{align}
using the expression of $\sigma$ in Eq.(\ref{eq:10}) or using the relation $\Vec{J}=\sigma\Vec{E}$, as far as the second order of $E$ is concerned.
The time dependence of temperature $T$ comes from the order $E^2$ and consequently this correction becomes of the order $E^4$ in Eq.(\ref{eq:33}), as is discussed below.

Now, the time dependence of $T(t)$ is effectively defined by the equation
\begin{equation}
U(t)=\langle \mathcal{H}_0\rangle _0(T(t))\equiv \text{Tr} \mathcal{H}_0\exp\left(-\beta(t)\mathcal{H}_0\right)/Z_0(\beta(t)),\label{eq:34}
\end{equation}
with $\beta(t)=1/k_{\text{B}}T(t)$, and $Z_0(\beta)=\text{Tr}\exp(-\beta\mathcal{H}_0)$, in a slowly varying state.
This is a convenient parameter to characterize "thermodynamically" non-equilibrium systems, which is expected to be observed experimentally with respect to the change of the internal energy.
Then, $T(t)$ is expanded in the form $T(t)=T+a(t)E^2+\cdots$.
Thus, the entropy production (\ref{eq:33}) is time-independent in the lowest order.

In general, according to the symmetry arguments on $\mathcal{H}_0$ and the current operator $\Vec{j}=\Vec{\dot{A}}$, we find that the average of the current, $\Vec{J}_E (t)=\langle \Vec{j}\rangle _t=\text{Tr}\Vec{j}\rho(t)$, is expressed by
\begin{equation}
\Vec{J}_E(t)=\text{Tr}\Vec{j}\rho_{\text{a}}(t),\label{eq:35}
\end{equation}
using the antisymmetric part $\rho_{\text{a}}(t)$ of $\rho(t)$ given by 
\begin{subequations}
\begin{equation}
\rho_{\text{a}}(t)=\frac{1}{2}\left\{ \rho(t;\Vec{E})-\rho(t;-\Vec{E}) \right\}\equiv P_{\text{antisym}}\rho(t)=\rho_1(t)+\rho_3(t)+\cdots,\label{eq:36a}
\end{equation}
and that the entropy production is given by $(dS/dt)_{\text{irr}}=\text{Tr}\mathcal{H}_0 \rho'_{\text{s}}(t)/T$ using the symmetric part
\begin{equation}
\rho_{\text{s}}(t)=\frac{1}{2}\left\{ \rho(t;\Vec{E})+\rho(t;-\Vec{E}) \right\}\equiv P_{\text{sym}}\rho(t)=\rho_0+\rho_2(t)+\cdots.\label{eq:36b}
\end{equation}
\end{subequations}
We note here that these partial density matrices satisfy the following "projected (symmetry-separated) von Neumann equations"
\begin{subequations}
\begin{equation}
i\hbar\frac{\partial}{\partial t}\rho_{\text{a}}(t)=[\mathcal{H}_0,\rho_{\text{a}}(t)]+[\mathcal{H}_1(t),\rho_{\text{s}}(t)],\label{eq:37a}
\end{equation}
and
\begin{equation}
i\hbar\frac{\partial}{\partial t}\rho_{\text{s}}(t)=[\mathcal{H}_0,\rho_{\text{s}}(t)]+[\mathcal{H}_1(t),\rho_{\text{a}}(t)].\label{eq:37b}
\end{equation}
\end{subequations}
These are written in the following unified form
\begin{equation}
i\hbar\frac{\partial}{\partial t}\rho_q(t)=[\mathcal{H}_0,\rho_q(t)]+[\mathcal{H}_1(t),(1-P_{q})\rho(t)],\label{eq:38}
\end{equation}
where $q$ denotes "sym" or "antisym" and $P_q$ is the projection operator to the "$q$" part of the density matrix $\rho(t)$; $P_q^2=P_q$ and $P_{\text{sym}}+P_{\text{antisym}}=1$.
This separation of symmetric and antisymmetric parts of $\rho(t)$ is shown to play an essential role in formulating steady states as is seen in the succeeding sections.

Using these partial (projected) density matrices, the general current $\Vec{J}_E$, conductivity $\sigma_E$ and entropy production $(dS/dt)_{\text{irr}}(E)$ are expressed, respectively, as
\begin{align}
\Vec{J}_E&=\text{Tr}\Vec{j}\rho_{\text{a}}(t)=\sigma_E\Vec{E},\label{eq:39}
\\
\sigma_E&=\sigma_0+\sigma_2E^2+\cdots+\sigma_{2n}E^{2n}+\cdots;\quad \sigma_0=\sigma,\label{eq:40}
\end{align}
and
\begin{align}
\left(\frac{dS}{dt} \right)_{\!\!\!\text{irr}}\!\!\!\!(E)&=\frac{1}{T(t)}\text{Tr}\mathcal{H}_0\rho'_{\text{sym}}(t)=\frac{1}{T(t)i\hbar}\text{Tr}\mathcal{H}_0[\mathcal{H}_1(t),\rho_{\text{a}}(t)]\notag
\\
&=\frac{1}{T(t)i\hbar}\text{Tr}[\mathcal{H}_0,\mathcal{H}_1(t)]\rho_{\text{a}}(t)=-\frac{1}{T(t)}\text{Tr}\dot{\mathcal{H}}_1(t)\rho_{\text{a}}(t)\notag
\\
&=\frac{\Vec{J}_E\cdot\Vec{E}}{T(t)}=\frac{\sigma_E E^2}{T(t)}=\frac{(\sigma_0+\sigma_2 E^2+\cdots)E^2}{T(t)}>0\label{eq:41}
\end{align}
for an arbitrary strength of the electric field $\Vec{E}$, as far as the entropy change is defined by $(dS/dt)_{\text{irr}}(E)=\text{Tr}\mathcal{H}_0\rho'(t)/T(t)$ in a non-stationary state, and as far as the definition of $T(t)$ in Eq.(\ref{eq:34}) is justified in a slowly-varying region.

Thus, the entropy production associated even with the linear transport phenomena is given by the time derivative of the {\itshape{second-order}} density matrix, $\rho'_2(t)$.
That is, the linear response scheme is closely related to the higher-order terms of the density matrix, when we discuss the relation between transport phenomena and the irreversibility associated with them.

As is seen from Eqs.(\ref{eq:39}) and (\ref{eq:41}), the current and entropy production are dual through the conductivity whose finiteness ($\sigma_E>0$) is the essence of irreversibility.

\section{New schemes of steady states using relaxation-type von Neumann equations}

In the preceding section, we have discussed a non-stationary state in order to clarify the entropy production of the relevant system described by the Hamiltonian $\mathcal{H}(t)=\mathcal{H}_0+\mathcal{H}_1$ with a static electric field $E$; $\mathcal{H}_1=-\Vec{A}\cdot\Vec{E}$.
In fact, the electric power (given by $\langle  \mathcal{H}_1 \rangle $) is transformed into the internal energy $U(t)=\langle \mathcal{H}_0\rangle _t$ (or heat-energy) inside the system described by $\mathcal{H}_0$ (even without interactions of it with the heat bath).
Thus, the entropy production is derived from the Hamiltonian $\mathcal{H}_0$ in the thermodynamic limit and then in the limit $t_0\rightarrow -\infty$ (corresponding to the coarse graining of time) with an adiabatic factor $e^{-\epsilon t}(\epsilon >0)$ if necessary.

In the present section, we study the entropy production in steady states corresponding to more realistic situations.
For this purpose, we have to introduce explicitly the interaction of the system with the heat bath, in order to extract generated heat outside of the system.
For this purpose, we extend the von Neumann equation (\ref{eq:1}) as follows:
\begin{equation}
\frac{\partial \rho(t)}{\partial t}=\frac{1}{i\hbar}[\mathcal{H}(t),\rho(t)]-\Gamma(\rho(t)),\label{eq:42}
\end{equation}
where the relaxation term $\Gamma(\rho(t))$ may be given in the Lindblad form\cite{18}:
\begin{equation}
\Gamma(\rho)=\sum_{\alpha}\left(V_{\alpha} \rho V_{\alpha}^{\dagger}-\frac{1}{2}\left\{V_{\alpha}^{\dagger}V_{\alpha},\rho\right\}\right).\label{eq:43}
\end{equation}
However, we are now interested only in the conceptual mechanism of the entropy production in steady states.
Thus, we make use of the simplest form of $\Gamma$, under the consideration of the essential difference between the symmetric density matrix $\rho_{\text{s}}(t)$ and antisymmetric one $\rho_{\text{a}}(t)$ introduced in the preceding section.
It should be remarked that the current is related to the conservation of particles or charges, and consequently that it is expressed easily by a kind of non-equilibrium density matrix including current operators \cite{7,11,19,20}.
Only  the adiabatic factor $e^{-\epsilon t}$ may be necessary for the convergence of the integral of the time-correlation function of the relevant current, which is described by $\rho_{\text{a}}(t)$.
Thus, we assume, for $\Gamma$,
\begin{equation}
\Gamma\rho_{\text{a}}(t)=\epsilon\rho_{\text{a}}(t)\quad(\epsilon\rightarrow +0).\label{eq:44}
\end{equation}

Although the total energy is a conserved quantity, the entropy (or heat) is not so.
The entropy characterizes a qualitative property of the energy in non-equilibrium states.
(Of course, it is more strictly defined as a thermodynamical state quantity in an equilibrium system.)
Therefore, we have to treat the heat change in a quite different way from the current.
The generated heat of the system should be extracted outside at a finite speed to construct a steady state and it is expressed by the change of $\rho_{\text{s}}(t)$.
Thus, we assume
\begin{equation}
\Gamma\rho_{\text{s}}(t)=\epsilon_{\text{r}}(\rho_{\text{s}}(t)-\rho_0);\quad\epsilon_{\text{r}}=\frac{1}{\tau_{\text{r}}}.\label{eq:45}
\end{equation}
The relaxation time $\tau_{\text{r}}$ in Eq.(\ref{eq:45}) is macroscopic, namely much larger than the collision time $\tau$ of electrons colliding with impurities or phonons (i.e., $\tau\ll\tau_{\text{r}}$), See Appendices A and B, concerning the physical role of the collision time $\tau$.

More explicitly we may write in the following:
\begin{equation}
\frac{d}{dt}\rho_{\text{a}}(t)=\frac{1}{i\hbar}[\mathcal{H}_0,\rho_{\text{a}}(t)]+\frac{1}{i\hbar}[\mathcal{H}_1,\rho_{\text{s}}(t)]-\epsilon\rho_{\text{a}}(t),\label{eq:46}
\end{equation}
and
\begin{equation}
\frac{d}{dt}\rho_{\text{s}}(t)=\frac{1}{i\hbar}[\mathcal{H}_0,\rho_{\text{s}}(t)]+\frac{1}{i\hbar}[\mathcal{H}_1,\rho_{\text{a}}(t)]-\epsilon_{\text{r}}\left( \rho_{\text{s}}(t)-\rho_0 \right).\label{eq:47}
\end{equation}
In a unified way, we have Eq.(\ref{eq:42}) with $\Gamma\rho(t)=\Gamma\left(\rho_{\text{a}}(t)\right)+\Gamma\left(\rho_{\text{s}}(t)\right)$.
The trace conservation of the density matrix $\rho(t)=\rho_{\text{s}}(t)+\rho_{\text{a}}(t)$ and its positivity are easily proved, as in the case of the Lindblad form (\ref{eq:43}).
Thus, we obtain the differential equation of the internal energy $U(t)$ for $\mathcal{H}_1=-\Vec{A}\cdot\Vec{E}$ as 
\begin{equation}
\frac{d}{dt}U(t)=\Vec{J}_E(t)\cdot\Vec{E}-\epsilon_{\text{r}}\left(U(t)-U(t_0)\right),\label{eq:48}
\end{equation}
using the relation
\begin{equation}
\frac{1}{i\hbar}\text{Tr}\mathcal{H}_0[\mathcal{H}_1,\rho_{\text{a}}(t)]=-\text{Tr}\dot{\mathcal{H}}_1\rho_{\text{a}}(t)=\text{Tr}(\Vec{j}\cdot\Vec{E})\rho_{\text{a}}(t)\equiv \Vec{J}_E(t)\cdot\Vec{E}. \label{eq:49}
\end{equation}
The solution of Eq.(\ref{eq:48}) is given by
\begin{equation}
U(t)=\int_{t_0}^{t}e^{-\epsilon_{\text{r}}(t-s)}\Vec{J}_E(s)\cdot\Vec{E}ds+U(t_0).\label{eq:50}
\end{equation}
From Eqs.(\ref{eq:46}) and (\ref{eq:47}) in the limit $t\rightarrow \infty$, we obtain the stationary density matrices $\rho_{\text{s}}^{\text{(st)}}$ and $\rho_{\text{a}}^{\text{(st)}}$:
\begin{equation}
\rho_{\text{s}}^{\text{(st)}}=(\Vec{1}-\mathcal{A}^{-1}\mathcal{B}\mathcal{C}^{-1}\mathcal{D})^{-1}\rho_0,\text{ and }\rho_{\text{a}}^{\text{(st)}}=\mathcal{C}^{-1}\mathcal{D}(\Vec{1}-\mathcal{A}^{-1}\mathcal{B}\mathcal{C}^{-1}\mathcal{D})^{-1}\rho_0.\label{eq:51}
\end{equation}
and thereby we can evaluate the stationary values $\Vec{J}_E^{\text{(st)}}$ and $U^{\text{(st)}}$.
Here, the hyper-operators $\mathcal{A},\mathcal{B},\mathcal{C}$ and $\mathcal{D}$ are defined by
\begin{equation}
\mathcal{A}=1-\omega_{\text{r}}\delta_{\mathcal{H}_0},\mathcal{B}=\omega_{\text{r}}\delta_{\mathcal{H}_1},\mathcal{C}=1-\omega_{\epsilon}\delta_{\mathcal{H}_0},\text{ and }\mathcal{D}=\omega_{\epsilon}\delta_{\mathcal{H}_1},\label{eq:52}
\end{equation}
where $\omega_{\epsilon}=1/(i\hbar\epsilon), \omega_{\text{r}}=1/(i\hbar\epsilon_{\text{r}})=\tau_{\text{r}}/(i\hbar)$ and the inner derivation $\delta_{Q}$ is defined by\cite{15}
\begin{equation}
\delta_{Q}R=[Q,R]=QR-RQ.\label{eq:53}
\end{equation}
The inverse hyper-operators $\mathcal{A}^{-1}$ and $\mathcal{C}^{-1}$ are defined in a power series as
\begin{equation}
\mathcal{A}^{-1}=(\Vec{1}-\omega_{\text{r}}\delta_{\mathcal{H}_0})^{-1}=\Vec{1}+\sum_{n=1}^{\infty}\left(\omega_{\text{r}}\delta_{\mathcal{H}_0}\right)^n \text{  etc.}\label{eq:54}
\end{equation}
For example, we have
\begin{align}
\rho_1^{\text{(st)}}&=\frac{1}{\Vec{1}-\omega_{\epsilon}\delta_{\mathcal{H}_0}}\omega_{\epsilon}\delta_{\mathcal{H}_1}\rho_0=\int_0^{\infty}dt \exp \left(\frac{t}{i\hbar}\delta_{\mathcal{H}_0} \right)\left(\frac{1}{i\hbar}[\mathcal{H}_1,\rho_0]\right) e^{-\epsilon t} \notag
\\
&=\rho_0\int_0^{\infty}dt\int_0^{\beta}d\lambda \Vec{j}(-t-i\hbar\lambda)\cdot\Vec{E}e^{-\epsilon t},\label{eq:55}
\end{align}
and
\begin{equation}
\rho_2^{\text{(st)}}=\int_0^{\infty}dt\exp \left( -(\epsilon_{\text{r}}-\frac{1}{i\hbar}\delta_{\mathcal{H}_0})t \right)\left(\frac{1}{i\hbar}[\mathcal{H}_1,\rho_1^{\text{(st)}}]\right). \label{eq:56}
\end{equation}

Explicit applications of these results to the entropy production will be presented in the next section.
(Note that $\mathcal{B}^{-1}$ and $\mathcal{D}^{-1}$ can not be defined.)

\section{Dynamical-derivative representation method to reveal the irreversibility and entropy production of steady states}

In the preceding section, we have made a formulation of the steady states $\rho_{\text{s}}^{\text{(st)}}$ and $\rho_{\text{a}}^{\text{(st)}}$.
Then, the averages of the current $\Vec{J}_E^{\text{(st)}}$ and the internal energy $U^{\text{(st)}}$ are given, respectively, by
\begin{equation}
\Vec{J}_E^{\text{(st)}}=\text{Tr}\Vec{j}\rho_{\text{a}}^{\text{(st)}} \quad {\text{ and }} \quad U^{\text{(st)}}=\text{Tr}\mathcal{H}_0\rho_{\text{s}}^{\text{(st)}}.\label{eq:57}
\end{equation}
Correspondingly, the stationary temperature $T_{\text{st}}$ is expressed using the relation
\begin{equation}
\langle \mathcal{H}_0\rangle _0(T_{\text{st}})=U^{\text{(st)}}=U^{\text{(st)}}(\tau_{\text{r}}),\label{eq:58}
\end{equation}
as a function of the relaxation time $\tau_{\text{r}}$.
More explicitly, from Eq.(\ref{eq:48}), we obtain the relation
\begin{equation}
\langle \mathcal{H}_0\rangle _0(T_{\text{st}})-\langle \mathcal{H}_0\rangle _0(T)=U^{\text{(st)}}-U(t_0)=\tau_{\text{r}}\Vec{J}_E^{\text{(st)}}\cdot\Vec{E}\quad\left(\Leftarrow \text{Tr}\Gamma^{-1}(\Vec{j}\rho_{\text{st}})\cdot\Vec{E} \right).\label{eq:59}
\end{equation}
This relation is also confirmed using Eqs.(\ref{eq:55}) and (\ref{eq:56}) up to the second order in $E$ and it gives the expression of $T_{\text{st}}$ as $T_{\text{st}}=T+\tau_{\text{r}}\sigma E^2/C(T)+O(E^4)$ with the specific heat $C(T)$ of the system described by $\mathcal{H}_0$.
Namely, the change of the stationary internal energy is given by the nonlinear Joule heat per unit time multiplied by the relaxation time $\tau_{\text{r}}$, as is expected physically.
Note that $T_{\text{st}}=T$ only when $\tau_{\text{r}}=0$.
This condition $(\tau_{\text{r}}=0)$ corresponds to the instantaneous extraction of generated heat outside of the system.
This is practically impossible.
We have $T_{\text{st}}>T$ (i.e., $\tau_{\text{r}}>0$) in a realistic situation.

Next, we study the entropy change $(dS/dt)_{\text{irr}}$ in the stationary state.
At a glance, it seems difficult to calculate from the time change of the internal energy (\ref{eq:48}), because the $dU(t)/dt$ goes to zero in the steady state.
In order to overcome this difficulty, we study the right-hand side of Eq.(\ref{eq:48}) from a physical point of view.
The first term $\Vec{J}_E(t)\cdot\Vec{E}$ of it is expressed in the form
\begin{equation}
\Vec{J}_E(t)\cdot\Vec{E}=\text{Tr}\mathcal{H}_0\Dd(\rho(t))=\text{Tr}(\tilde{\Dd}\mathcal{H}_0)\rho(t),\label{eq:60}
\end{equation}
where
\begin{equation}
\tilde{\Dd}\mathcal{H}_0\equiv \frac{1}{i\hbar}[\mathcal{H}_0,\mathcal{H}(t)],\text{ and }\Dd(\rho(t))\equiv \frac{1}{i\hbar}[\mathcal{H}(t),\rho(t)]=\frac{\partial \rho(t)}{\partial t}+\Gamma(\rho(t)), \label{eq:61}
\end{equation}
as is easily seen from the derivation of Eq.(\ref{eq:48}) from Eq.(\ref{eq:47}).
Here, the hyper-operators $\Dd$ and $\tilde{\Dd}$ denote the time derivatives of the density matrix and an operator, respectively, in the above sense (\ref{eq:61}).
We call these as "dynamical-derivative representations", which may correspond to an intermediate representation between the Heisenberg and Schr$\ddot{\text{o}}$dinger representations.
Note that the hyper-operator $\tilde{\Dd}$ defined by $(i/\hbar)\delta_{\mathcal{H}(t)}$ satisfies the Leibniz rule $\tilde{\Dd}(AB)=(\tilde{\Dd}(A))B+A(\tilde{\Dd}(B))$, similarly to ordinary differential operators.
We remark here that $\tilde{\Dd}(A)$ is generally different from the time derivative $\dot{A}$ defined by $\dot{A}=(i\hbar)^{-1}\delta_{\mathcal{H}_0}A$, but it happens that $\tilde{\Dd}(\mathcal{H}_1)=\dot{\mathcal{H}}_1$ for $\mathcal{H}=\mathcal{H}_0+\mathcal{H}_1$.

\begin{figure}
\begin{center}
\includegraphics[width=7cm,clip]{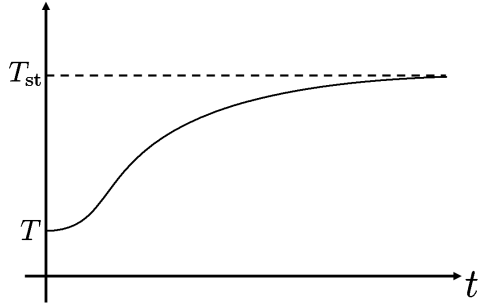}
\end{center}
\caption{Schematic figure of the time dependence of the temperature $T(t)$ of the system.}
\label{fig1}
\end{figure}

The mechanism of heat generation should be the same as in the absence of the relaxation term $\Gamma\rho(t)$.
Consequently, the entropy production in our new situation described by Eq.(\ref{eq:47}) is expressed in the form
\begin{equation}
\left(\frac{dS}{dt}\right)_{\!\!\!\text{irr}}\!\!(t)=\text{Tr}S\Dd\rho(t)=\frac{1}{T(t)}\text{Tr}(\tilde{\Dd}(\mathcal{H}_0))\rho(t)=\left(\Vec{J}_E(t)\cdot\Vec{E}\right)/T(t),\label{eq:62}
\end{equation} 
with the time-dependent temperature $T(t)$ defined by Eq.(\ref{eq:34}), namely
\begin{equation}
\langle \mathcal{H}_0\rangle _0(T(t))=U(t)\equiv \langle \mathcal{H}_0 \rangle _t\equiv \text{Tr}\mathcal{H}_0\rho(t),\label{eq:63}
\end{equation}
together with Eq.(\ref{eq:50}).
Thus, the temperature $T(t)$ defined by Eq.(\ref{eq:63}) behaves like Fig.\ref{fig1}.
In order to emphasize the logic of our theory, we rewrite Eq.(\ref{eq:48}) in the following form
\begin{equation}
\frac{dU(t)}{dt}=\bigg( \text{Tr} (\tilde{\Dd}(\mathcal{H}_0)) \rho(t)\bigg)_{\text{generated heat}}\!\!\!\!-\quad\bigg(\frac{U(t)-U(t_0)}{\tau_{\text{r}}}\bigg)_{\text{extracted heat}}\label{eq:64}
\end{equation}
which vanishes in the steady state, owing to the balance of the two terms in the right-hand side of Eq.(\ref{eq:64}).
It should also be emphasized that the extraction of heat energy in Eq.(\ref{eq:64}) is given by the symmetric part $\rho_{\text{sym}}(t)$ through $(\rho_{\text{sym}}(t)-\rho_0)/\tau_{\text{r}}$.

\section{Summary and discussion}

In the present paper, we have succeed in deriving, from first principles, the entropy production in transport phenomena for a typical example of static electric conduction.
This yields a direct derivation of irreversibility in transport phenomena.
It is very interesting that the {\itshape{second-order}} term $\rho_2(t)$ of the density matrix gives the lowest-order contribution to the entropy production.
Of course, it is expressed finally by the first-order $\rho_1(t)$ produced by $\dot{\mathcal{H}}_1(t)$, and any higher-order contribution can be expressed by the multiple integrals of commutators of $\rho_1(t)$ with $\{\mathcal{H}_1(t)\}$, as shown in Eq.(\ref{eq:30}).
After the discovery of the present derivation of the entropy production in transport phenomena, it seems to be quite natural, because the entropy production is a symmetric (and consequently even) function of an external force $\Vec{F}$ (or $\Vec{E}$ in the case of electric conduction) and consequently because the lowest-order contribution should start from the second order in $\Vec{F}$ (or $\Vec{E}$), as was already discussed phenomenologically by many authors (see Section 3).

In any way, the present discovery of the appearance of entropy production in the second-order term of the density-matrix expansion is one of the very rare cases in which the essence of each physical phenomenon appears in the second-order term of each perturbational expansion, as in the Kondo effect\cite{21} and in the negative divergence of the nonlinear susceptibility of spin glasses at the transition point\cite{22}.

It is also instructive to remark that the irreversibility and entropy production are based on the existence of the non-vanishing and finite transport coefficients (such as the conductivity $\sigma$), as has been already briefly mentioned in Section 2 concerning the order of taking the two limiting procedures the volume $V\rightarrow \infty$ and the initial time $t_0\rightarrow -\infty$.
This yields the statement that the irreversibility is assured by the inequality $\Delta_{\text{s}}<\Delta_{\text{t}}$ of the two scaling exponents $\Delta_{\text{s}}$ and $\Delta_{\text{t}}$ of the space scaling $x'=x/\epsilon^{\Delta_{\text{s}}}$ and time scaling $t'=t/\epsilon^{\Delta_{\text{t}}}$ for small $\epsilon$ $(\epsilon>0)$, concerning the equation governing the relevant intrinsically irreversible phenomenon.
For a diffusion process (a typical irreversibility process), for example, we have $\Delta_{\text{t}}=2\Delta_{\text{s}}$, because $(\Delta x)^2\sim(\Delta t)$ in this case.
Thus, space and time have quite different characters in the theory of probability (or stochasticity) and entropy production.
Note also that the above condition $\Delta_{\text{s}}<\Delta_{\text{t}}$ is satisfied in difference equations in which chaotic phenomena (and consequently irreversible phenomena) are apt to occur.
When the relevant current operator contains partially a constant of motion, there appears the problem of ergodicity concerning linear responses and possible constants of motion\cite{23}.

By the way, it may be interesting to remark that 
\begin{equation}
\text{Tr}\mathcal{H}_1(t)\rho'_{2n-1}(t)=-\text{Tr}\mathcal{H}_0\rho'_{2n}(t)=\text{Tr}\dot{\mathcal{H}}_1(t)\rho_{2n-1}(t),\label{eq:65}
\end{equation}
for all $n$, and consequently that
\begin{align}
\text{Tr}\mathcal{H}_1(t)\rho'_{\text{a}}(t)&=-\text{Tr}\mathcal{H}_0\rho'_{\text{s}}(t)=\text{Tr}\dot{\mathcal{H}}_1(t)\rho_{\text{a}}(t)=\langle \dot{\mathcal{H}}_1(t)\rangle _t\notag
\\
&=-\frac{d}{dt}\langle \mathcal{H}_0\rangle  _t=-\langle \mathcal{H}_0\rangle '_t;\quad \rho'(t)\equiv \frac{d}{dt} \rho(t).\label{eq:66}
\end{align}
These relations yield again the proposition that the time derivative of the relative entropy (\ref{eq:23}) defined by the total Hamiltonian always vanishes, if the contribution of the second-order term is also included, as was briefly mentioned in subsection (3.4).

Physically speaking, the quantity $\langle  \dot{\mathcal{H}}_1(t) \rangle _t=\text{Tr}\mathcal{H}_1(t)\rho'(t) (<0)$ denotes the loss of electric power inside the relevant system, which is implicitly assumed to be compensated from outside, in order to keep the electric current.
To understand this situation more clearly, the waterfall analogy will be helpful, as is often used.
The gravitational energy of water ($\rightarrow $electric field effect) is transformed into the kinetic energy of the water changing into heat ($\rightarrow $internal energy) at the bottom of the waterfall and consequently the temperature there increases.
However, at the top of the fall, water should be supplied (from a river or a lake) to keep the stream of the waterfall.

As has often been explained from the energy balance, the reader might insist that the loss of electric power,$\langle \dot{\mathcal{H}}_1\rangle _t(=-\Vec{J}\cdot\Vec{E})$ should express the energy dissipation and that it can be calculated by $\rho_1(t)$.
However, logically speaking, the loss of electric power might change to mechanical energy, which does not necessarily mean immediately the generation of heat energy.
In fact, many people including the author have not been satisfied with the argument due to the above energy balance.
The essence of irreversibility should be explained directly by the generation of heat inside the system in the case of electric conduction.
The change of the internal energy $\langle \mathcal{H}_0\rangle _t$ corresponding to the heat energy is logically derived from the time derivative of the second-order (symmetric) term $\rho_2(t)$ of the density matrix, and consequently that it is finally equivalent to the quantity $-\langle \dot{\mathcal{H}}_1\rangle _t(>0)$.

The reader might say that the relation $\langle \mathcal{H}_0\rangle '_t=-\langle \dot{\mathcal{H}}_1 \rangle _t$ is nothing but the energy conservation.
It is true in the isolated system (\ref{eq:1}).
However, it comes from the exact unitary evolution of the density matrix $\rho(t)$.
To prove this explicitly requires the calculation of $\langle \mathcal{H}_0+\mathcal{H}_1\rangle _t$ up to the second order of an external field, using $\rho_0+\rho_1(t)+\rho_2(t)$, even in the lowest-order argument, because $\text{Tr}\mathcal{H}_0\rho_1(t)=0$ from the symmetry argument.
Anyway, we have to use the symmetric part of $\rho(t)$ in order to discuss the energy change and entropy production.
Thus, the present direct derivation of the internal (heat) energy using the symmetric part $\rho_{\text{sym}}(t)$ manifests the irreversibility of the electric conduction.
This observation has inspired the author to find the symmetry-separated relaxation-type von Neumann equation.
This symmetry-separated treatment of $\rho(t)$ is crucial in formulating non-equilibrium steady states with entropy production, as has been presented in Section 6.
The present theory insists that the internal energy is derived logically from the second-order or higher-order symmetric density matrix $\rho_{\text{sym}}(t)$.

In our system of electric conduction (or any other transport phenomenon), we assume that our system is infinite to avoid the problem of boundary conditions and to assure the irreversibility of these processes.

It should be emphasized again that our argument on the irreversibility and entropy production has become remarkably transparent by confining our explicit analysis in a typical example of static electric conduction and furthermore by treating first a non-stationary state and then as simple steady states as possible.
Such a stationary density matrix formalism as gives the entropy production has been constructed in the present paper.
More explicitly, $\rho_1^{\text{(st)}}$ agrees with Kubo's expression (Eq.(\ref{eq:55})).
However, $\{\rho_n^{\text{(st)}}(n\geq 2)\}$ contain the parameter $\tau_{\text{r}}$ which controls the speed to extract heat outside.

Thus, the two research trends of Prigogine's non-equilibrium thermodynamics based on the entropy production (irreversibility) and Kubo's linear response framework are now integrated by clarifying the role of symmetry of non-equilibrium states (i.e., $\rho_{\text{a}}(t)$ for current and $\rho_{\text{s}}(t)$ for irreversibility, showing duality).

The present theory\cite{24} can be easily extended to more general transport phenomena (including more complicated non-linear transport phenomena\cite{25}) in non-stationary and steady states both with electric and heat currents, in which Onsager's reciprocity relations and the KMS condition will be extended as will be reported in the near future.

\section*{Acknowledgements}

The author would like to didicate the present paper to his wife Noriko who passed away during the study of the present theory.
The author would like to thank the referees for valuable comments and Y.Hashizume for discussion and digitization of the present manuscript.
This work was supported by the Grant-in-Aid for Science Research on Priority Area "Deepening and Expansion of Statistical Mechanical Informatics".

\appendix
\section{Derivation of entropy production using the relaxation-type Boltzmann equation}

As has been explained in detail in Sections 4,5 and 6, the generation of heat or entropy production occurs inside the system described by $\mathcal{H}_0$.
In order to illustrate this situation more intuitively, we discuss the electric current and entropy production using the following "relaxation-type Boltzmann equation"\cite{7}
\begin{equation}
\frac{\partial f(t)}{\partial t}+\Vec{v}\cdot\frac{\partial f(t)}{\partial \Vec{x}}+e\Vec{E}\cdot\frac{\partial f(t)}{\partial\Vec{p}}=-\frac{1}{\tau}(f(t)-f_0)\label{eq:a1}
\end{equation}
for the distribution function of electrons, $f(t)$.
In the present uniform case, the second term of the left-hand side of Eq.(\ref{eq:a1}) vanishes.
We expand $f(t)$ as
\begin{equation}
f(t)=f_0+f_1(t)+f_2(t)+\cdots.\label{eq:a2}
\end{equation}
If we note the relation $\epsilon=p^2/2m$, the third term of the left-hand side of Eq.(\ref{eq:a1}) is expressed in the form
\begin{equation}
e\Vec{E}\cdot\frac{\partial f(t)}{\partial \Vec{p}}=e\Vec{v}\cdot\Vec{E}\frac{\partial f}{\partial \epsilon}.\label{eq:a3}
\end{equation}
The first-order equation of Eq.(\ref{eq:a1}) is given by
\begin{equation}
f_1(\epsilon)^{\text{(st)}}=-\tau e\Vec{v}\cdot\Vec{E}\frac{\partial f_0}{\partial \epsilon}=\tau(\Vec{j}\cdot\Vec{E})\beta f_0\label{eq:a4}
\end{equation}
in the steady state.
Then the stationary current $\Vec{J}^{\text{(st)}}$ is given by
\begin{equation}
\Vec{J}^{\text{(st)}}=\int e\Vec{v}f_1^{\text{(st)}}(\epsilon) D(\epsilon) d\epsilon =\frac{ne^2\tau}{m}\Vec{E}=\sigma \Vec{E},\label{eq:a5}
\end{equation}
as is well known.
Here, $D(\epsilon)$ denotes the density of states, and $\tau$ is the collision time of electrons, which should be calculated, in principle, from the Hamiltonian $\mathcal{H}_0$.

The second-order equation of Eq.(\ref{eq:a1}) is given in the form 
\begin{equation}
\frac{\partial f_2}{\partial t}+e\Vec{v}\cdot\Vec{E}\frac{\partial f_1}{\partial \epsilon}=-\frac{1}{\tau}f_2.\label{eq:a6}
\end{equation}
Then, we obtain
\begin{equation}
\left(f'_2(\epsilon)\right)_{\text{heat gen.}}=\frac{1}{\tau}f_2^{\text{(st)}}(\epsilon)=-(\Vec{j}\cdot\Vec{E})\frac{df_1^{\text{(st)}}(\epsilon)}{d\epsilon}\simeq \beta(\Vec{j}\cdot\Vec{E})f_1^{\text{(st)}}.\label{eq:a7}
\end{equation}
It is easily shown that the entropy production defined by
\begin{equation}
\left(\frac{dS}{dt}\right)_{\!\!\!\text{irr}}=\frac{1}{T}\int \epsilon \bigg(f'_2(\epsilon,t)\bigg)_{\!\!\!\text{heat gen.}}\!\!\!\!\!\!\!\!\!\!\!\!\!\!\!\!D(\epsilon)d\epsilon,\label{eq:a8}
\end{equation}
is given by
\begin{equation}
\left(\frac{dS}{dt}\right)_{\!\!\!\text{irr}} \simeq \frac{1}{T}\int\beta\epsilon (\Vec{j}\cdot\Vec{E})f_1^{\text{(st)}}(\epsilon)D(\epsilon)d\epsilon \simeq \frac{1}{T}(\Vec{J}^{\text{(st)}}\cdot\Vec{E})=\frac{\sigma E^2}{T},\label{eq:a9}
\end{equation}
which agrees approximately with the expression obtained in the text.
The present argument yields again our statement that the entropy production is derived from the second-order term of the distribution function or the density matrix.

\section{Einstein-Onsager Ansatz and Relaxation-time Approximation of $\sigma$ in the Kubo formula}

It may be instructive to rederive Eq.(\ref{eq:a5}) from Eq.(\ref{eq:10}) using Einstein's theory of Brownian motion\cite{26}:
\begin{equation}
m\frac{d\Vec{v}_j}{dt}=-\zeta\Vec{v}_j+\Vec{\eta}_j(t)+\Vec{F}\label{eq:b1}
\end{equation}
for the velocity $\Vec{v}_j$ of the $j$-th particle with the Gaussian white noise $\Vec{\eta}_j(t)$ satisfying the relation
\begin{equation}
\langle \Vec{\eta}_j(t)\Vec{\eta}_j(t')\rangle =2\epsilon'\delta(t-t').\label{eq:b2}
\end{equation}
Here, $\Vec{F}$ denotes an external force and we assume $\Vec{F}=e\Vec{E}$ in our static electric conduction.
If we take the average of Eq.(\ref{eq:b1}) and take the summation of the whole electrons in unit volume, then we obtain
\begin{equation}
\frac{d}{dt}\Vec{J}(t)=-\gamma \Vec{J}(t)+\left(\frac{ne^2}{m}\right)\Vec{E};\quad \gamma=\frac{\zeta}{m}\label{eq:b3}
\end{equation}
Thus, the macroscopic relaxation law is the same as the microscopic one (\ref{eq:b1}) except the noise $\Vec{\eta}_j(t)$ in the linear regime.
Physically speaking, this should be stated inversely.
That is, the introduction of the systematic force $-\zeta\Vec{v}_j$ (whose coefficient is identified as the macroscopic one) in Eq.(\ref{eq:b1}) is due to Einstein's drastic intuition.
This idea was also effectively used by Onsager\cite{13} in deriving the reciprocity theorem.
This may be called Einstein-Onsager ansatz.
In the steady state, we arrive at
\begin{equation}
\Vec{J}^{\text{(st)}}=\sigma \Vec{E};\quad\sigma=\frac{ne^2}{m\gamma}=\frac{ne^2}{m}\tau,\quad\tau=\frac{1}{\gamma}.\label{eq:b4}
\end{equation}
This agrees with Eq.(\ref{eq:a5}), as it should be.
Similarly we obtain the current-current correlation
\begin{equation}
\frac{d}{dt}\langle \Vec{j}(t)\Vec{j}(0)\rangle _0=-\gamma\langle \Vec{j}(t)\Vec{j}(0)\rangle _0\label{eq:b5}
\end{equation}
from Eq.(\ref{eq:b1}) in the absence of an electric field.
The average $\langle \cdots\rangle _0$ denotes the thermal average for $\Vec{E}=0$ at the temperature $T$.
The solution of (\ref{eq:b5}) is given by
\begin{equation}
\langle \Vec{j}(t)\Vec{j}(0)\rangle _0=\langle \Vec{j}^2(0)\rangle _0 e^{-\gamma t}=\frac{ne^2k_{\text{B}}T}{m}e^{-\gamma t}.\label{eq:b6}
\end{equation}
This yields again Eq.(\ref{eq:a5}) using Eq.(\ref{eq:10}) in the classical limit $\hbar\rightarrow 0$.
By the way, the strength $\epsilon'$ of the noise is easily shown to be expressed as
\begin{equation}
\epsilon'=k_{\text{B}}T\zeta=k_{\text{B}}Tm\gamma\label{eq:b6}
\end{equation}
This is nothing but the classical fluctuation-dissipation relation\cite{7}.

\bibliographystyle{model1a-num-names}
\bibliography{<your-bib-database>}







\end{document}